\documentclass[aps,prl,twocolumn,preprintnumbers,superscriptaddress,10pt]{revtex4-1}
\usepackage{epsf,color,amsmath,amssymb,amsfonts}
\usepackage{enumerate}
\usepackage{hyperref}
\usepackage{graphicx}
\usepackage{psfrag}
\usepackage{dcolumn,bm}

\newcommand{\bea}{\begin{eqnarray}}
\newcommand{\beal}[1]{\begin{eqnarray}\label{#1}}
\newcommand{\eea}{\end{eqnarray}}
\newcommand{\be}{\begin{equation}}
\newcommand{\bel}[1]{\begin{equation}\label{#1}}
\newcommand{\ee}{\end{equation}}

\newcommand{\bit}{\begin{itemize}}
\newcommand{\eit}{\end{itemize}}
\newcommand{\ben}{\begin{enumerate}}
\newcommand{\een}{\end{enumerate}}

\newcommand{\mt}[1]{\textrm{\tiny #1}}
\newcommand{\nc}{N_\mt{c}}
\newcommand{\tiso}{t_\mt{iso}}

\newcommand{\ed}{{\cal E}}

\renewcommand{\pl}{{\cal P_\mt{L}}}
\newcommand{\pt}{{\cal P_\mt{T}}}
\newcommand{\deltap}{\Delta {\cal P}}
\newcommand{\sac}{\, , \qquad}

\newcommand{\eqn}[1]{(\ref{#1})}

\begin{document}
\onecolumngrid

\preprint{ICCUB-12-009}
\preprint{ITP-UU-12/04}

\title{Strong coupling isotropization of non-Abelian plasmas simplified}
\author{Michal P.~Heller}
\altaffiliation[On leave from: ]{\emph{National Centre for Nuclear Research,  Ho{\.z}a 69, 00-681 Warsaw, Poland.}}
\affiliation{Instituut voor Theoretische Fysica, Universiteit van Amsterdam \\
Science Park 904, 1090 GL Amsterdam, The Netherlands}
\author{David Mateos}
\affiliation{{Instituci\'o Catalana de Recerca i Estudis Avan\c cats (ICREA), 
Barcelona, Spain}}
\affiliation{{Departament de F\'\i sica Fonamental \&  Institut de Ci\`encies del Cosmos (ICC), Universitat de Barcelona, Mart\'{\i}  i Franqu\`es 1, E-08028 Barcelona, Spain}}
\author{Wilke van der Schee}
\affiliation{Institute for Theoretical Physics and Institute for Subatomic Physics,
Utrecht University, Leuvenlaan 4, 3584 CE Utrecht, The Netherlands}
\author{Diego Trancanelli}
\affiliation{\it Instituto de F\'\i sica, Universidade de S{\~a}o Paulo, 05314-970 S{\~a}o Paulo, Brazil}


\begin{abstract}
\noindent
We study the isotropization of a homogeneous, strongly coupled, non-Abelian plasma by means of its gravity dual. We compare the time evolution of a large number of initially anisotropic states as determined, on the one hand, by the full non-linear Einstein's equations and, on the other, by the Einstein's equations linearized around the final equilibrium state. The linear approximation works remarkably well even for states that exhibit large anisotropies. For example, it predicts with a 20\% accuracy the isotropization time, which is of the order of $\tiso \lesssim 1/T$, with $T$ the final equilibrium temperature. We comment on possible extensions to less symmetric  situations.

\end{abstract}
\maketitle

\noindent
{{\bf 1. Introduction.}}
Motivated by the strongly coupled nature of the quark-gluon plasma, much has been learned about the (near) equilibrium properties of strongly coupled plasmas by employing their dual description as a (slightly perturbed) static black hole (see \cite{review} and references therein).

The formation and far-from-equilibrium evolution of a plasma correspond on the gravity side to the formation of a far-from-equilibrium black hole and its subsequent relaxation. An outstanding problem in this context is understanding the short isotropization time of the quark-gluon plasma. In principle, studying this problem on the gravity side requires solving the full nonlinear Einstein's equations (EEQs), which typically can be done only numerically. 
Examples of such numerical studies include \cite{incl}-\cite{
Bantilan:2012vu}.



One of our purposes is to show that the problem on the gravity side can be simplified, at least in certain circumstances. Inspiration comes from the so-called `close limit approximation' (CLA) \cite{Price:1994pm} in the context of black hole mergers in four-dimensional  general relativity in asymptotically flat spacetime. The CLA is the statement that, once a single horizon forms around the two incident black holes, its subsequent evolution is well described by the EEQs linearized around the final equilibrium black hole, despite the fact that the initial horizon may or may not seem to be a small perturbation of the final one.
In particular, the form of the gravitational radiation emitted to infinity in the merger-plus-ring-down phase is well described by the CLA \cite{Anninos:1995vf}.

\begin{figure}[t!]
\centering{}
\includegraphics[width=9cm]{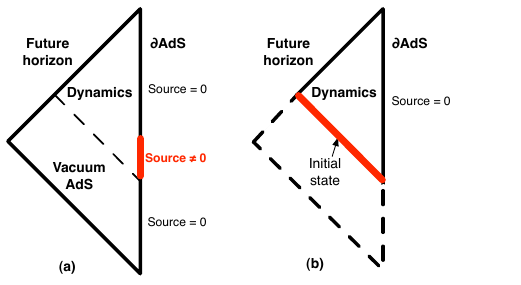}
\caption{(a) Setup of Ref.~\cite{Chesler:2008hg}. (b) Our setup. The initial state is specified on a $t=\mbox{const.}$ surface spanned by the radial coordinate $r$ --- see eqn.~\eqn{mansatz}. 
}
\label{penrose}
\end{figure}

Following \cite{Chesler:2008hg}, we study isotropization of a homogeneous plasma in a four-dimensional conformal field theory (CFT) in flat Minkowski space; on the gravity side this means that we work in the Poincar\'e patch of AdS$_5$. Note that, because of the homogeneity, the isotropization process involves exclusively non-hydrodynamic modes. Ref.~\cite{Chesler:2008hg} `creates' a far-from-equilibrium state by acting on the CFT vacuum with an external anisotropic source (see Fig.~\ref{penrose}).
In contrast, we study the isotropization of a large number of anisotropic initial  states in the absence of external sources (see \cite{Beuf:2009cx,Heller:2011ju} for related work). 
Each state is specified on the gravity side by an entire function on an initial-time slice (ITS), and hence it is characterized by an arbitrary number of scales. 

In this Letter we  focus on the time evolution of the expectation value of the gauge theory stress tensor; other observables will be considered elsewhere. Conservation of the stress tensor for a homogeneous plasma in the absence of external sources implies
that the energy density $\ed$ (but not the entropy density) must be constant in time. Consequently, the final equilibrium state is known without solving for the dynamical evolution. On the CFT side it is a homogeneous, isotropic plasma with temperature $\propto \ed^{1/4}$ and pressure $\ed/3$, with $\ed$ the initial energy density. On the gravity side it is a static, isotropic black brane with the same temperature. This makes the linear approximation (LA) particularly simple: we linearize EEQs around the static black brane and use them to evolve each initial state. As expected on general grounds, 
the dynamical evolution shows that an event horizon (but not necessarily an apparent horizon) is already present on the ITS for each of the states we consider. By comparing on the gravity side the full numerical evolution with the LA, we will see that the latter predicts with surprising accuracy the time evolution of the CFT stress tensor (see \cite{Bantilan:2012vu} for related observations), in analogy with the prediction of the gravitational radiation at infinity by the CLA. We emphasize that the applicability of the LA is not guaranteed a priori, since in general our initial states are not near-equilibrium states.


\noindent
{{\bf 2. Holographic model.}} For a conformal $SU(\nc)$ gauge theory, conservation and tracelessness of the stress tensor, together with homogeneity and rotational invariance in one plane (assumed for simplicity) imply that the stress tensor can be written as
\bea
&&\langle T_{\mu \nu} \rangle = \frac{\nc^2}{2\pi^2}
\mathrm{diag} \Big[ \ed, \, \pl(t), \, \pt(t), \, \pt(t) \Big] \,, \\
&&\pl(t) = \frac{1}{3} \ed - \frac{2}{3} \deltap (t) \,\,\,,\,\,\,
\pt (t) = \frac{1}{3} \ed + \frac{1}{3} \deltap (t)\,,\,\,\,\,\,\,\,\,
\eea
in terms of a single function $\deltap = \pt-\pl$ that measures the degree of anisotropy. Accordingly, the dual metric can be written as \cite{Chesler:2008hg}
\begin{equation}
\label{mansatz}
ds^2 = 2 dt dr - A dt^{2} + \Sigma^{2} e^{-2 B} dx_\mt{L}^{2} + \Sigma^{2}e^{B} d\mathrm{\mathbf{x}}_\mt{T}^{2} \,,
\end{equation}
where $A$, $\Sigma$ and $B$ are all functions of time $t$ and of the radial coordinate $r$. In the absence of CFT sources these are subject to the following boundary conditions near the AdS$_5$ boundary  at $r\to \infty$:
\begin{eqnarray}
\label{nearbdryexpansions}
&&A= r^2 + \frac{a_{4} }{r^{2}} - \frac{2b_{4}(t)^{2} }{7 r^{6}}  + \cdots \,, \cr
& &B=\frac{b_{4}(t)}{r^4} + \frac{b_{4}'(t)}{r^5}  +  \cdots
\,, \quad  \Sigma = r - \frac{b_{4}(t)^{2}}{7 r^{7}}  + \cdots \,. \,\,\,\,\,\,\,
\end{eqnarray}
As usual, the normalizable modes $a_4$ and $b_4(t)$ are not determined by the boundary conditions but must be read off from a full bulk solution that is regular in the interior. These modes are dual to the expectation value of the stress tensor. 
For the specific case of $SU(\nc)$ ${\cal N}=4$ super Yang-Mills theory this relation is  
\be
\ed = - 3 a_{4}/4 \quad \mathrm{and} \quad 
\deltap (t) = 3 b_{4}(t) \,.
\label{related}
\ee
Although $\ed$ is constant in time, a physical temperature can only be assigned to the system once (near) equilibrium is reached. In this regime  $\ed=3\pi^4 T^4/4$.

In the generalized Eddington-Finkelstein coordinates \eqn{mansatz} EEQs take the nested form
\begin{subequations}
\begin{eqnarray}
\label{Seq}
0 &=& \Sigma \, (\dot \Sigma)' + 2 \Sigma' \, \dot \Sigma - 2 \Sigma^2\,,
\\ \label{Beq}
0 &=& \Sigma \, (\dot B)' + {\textstyle \frac{3}{2}}
    \big ( \Sigma' \dot B + B' \, \dot \Sigma \big )\,,
\\  \label{Aeq}
0 &=& A'' + 3 B' \dot B - 12 \Sigma' \, \dot \Sigma/\Sigma^2 + 4\,,
\\  \label{Cr}
0 &= & \ddot \Sigma
    + {\textstyle \frac{1}{2}} \big( \dot B^2 \, \Sigma - A' \, \dot \Sigma \big)\,,
\\ \label{Ct}
0 &=& \Sigma'' + {\textstyle \frac{1}{2}} B'^2 \, \Sigma\,,
\end{eqnarray}
\label{Eeqns}%
\end{subequations}
where $h' \equiv \partial_r h$ and 
$\dot h \equiv \partial_t h + 
{\textstyle \frac{1}{2}} A \, \partial_r h$ are derivatives along ingoing and outgoing null geodesics, respectively. 
Eqs.~\eqn{Seq}-\eqn{Aeq} are dynamical equations, whereas eqs.~\eqref{Cr} and \eqref{Ct} are constraints. If the former hold, then \eqref{Cr} and \eqn{Ct} are satisfied everywhere provided they are satisfied near the AdS boundary and on the ITS, respectively. As is clear from the causal structure in Fig.~\ref{penrose}, the dynamical equations together with the constraints determine the solution in the region labeled `dynamics'. We find this solution by numerically evolving the full EEQs following the procedure outlined in \cite{Chesler:2008hg}.

Eqn.~\eqn{Ct} is a constraint on the possible initial states because it relates two of the metric functions on the ITS. We choose $B$ as the independent variable because it is directly related to the CFT anisotropy through eqn.~\eqn{related}. Thus each initial state is specified by a constant $a_4$ and a function of the radial coordinate $B(t=0,r)$. Note that for positive $\Sigma$ the constraint \eqref{Ct} implies $\Sigma''\leq 0$, which in combination with the asymptotic behavior $\Sigma \simeq r$ means that $\Sigma$ will vanish at some $r \geq0$ on the ITS. 
Preliminary explorations indicate that this is a curvature singularity, and we will come back to this issue elsewhere. In any case, for all the initial states which our numerical code was able to evolve in a stable manner, the region where $\Sigma=0$ was hidden behind an event horizon and hence it had no effect on the physics. 

The static black brane solution of \eqn{Eeqns} dual to a plasma in perfect equilibrium takes the form 
\be
A = r^2 (1- r_\mt{h}^ 4/r^{4}) \sac \Sigma = r \sac B=0 \,,
\label{blackbrane}
\ee
with the horizon located at $r_\mt{h}=\pi T$. Considering small fluctuations around these equilibrium values one finds that $A$ and $\Sigma$ are unmodified at linear order, whereas the $B$-fluctuation obeys eqn.~\eqn{Beq} with $\Sigma$ and $A$ as in \eqn{blackbrane}. 
 
\noindent
{{\bf 3. Results.}} We report on around 1000 initial states (for all of which our numerics converged nicely), most of them generated by a random procedure (to be explained in \cite{preparation}). For some profiles, an apparent horizon was present on the ITS, for some others it was not. On the one hand we determined the time evolution of each state by means of the full, non-linear EEQs. On the other hand we  solved the linear equation for $B$. In each case we read off the pressure anisotropy by extracting $b_4(t)$ from the near-boundary behavior \eqn{nearbdryexpansions}. 
\begin{figure*}
\begin{tabular}{cc}
\includegraphics[width=0.42 \textwidth]{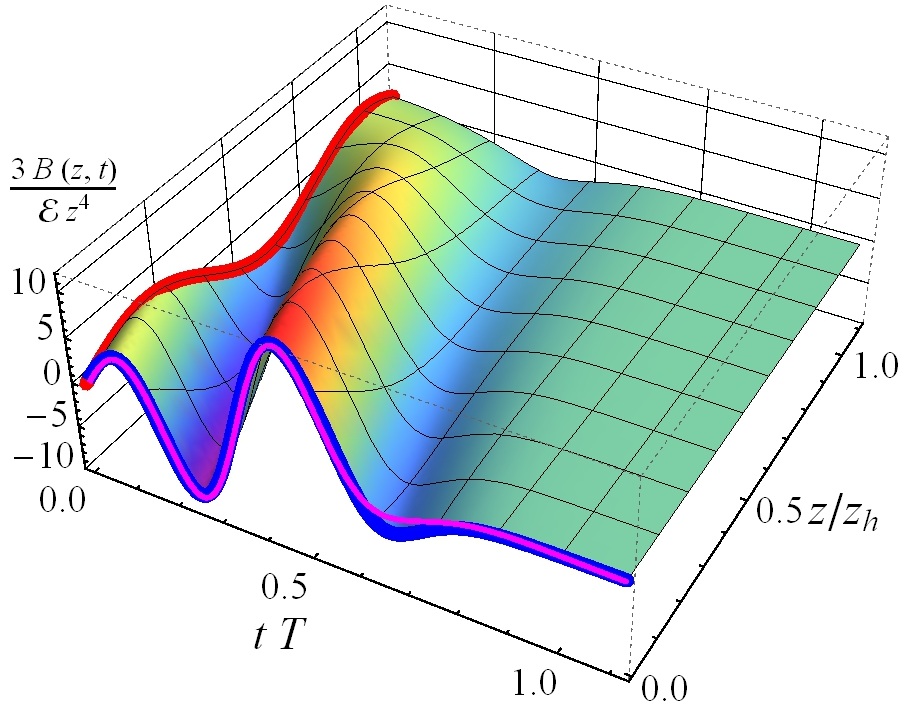} 
\quad\quad\quad & \quad\quad\quad
\includegraphics[width=0.42 \textwidth]{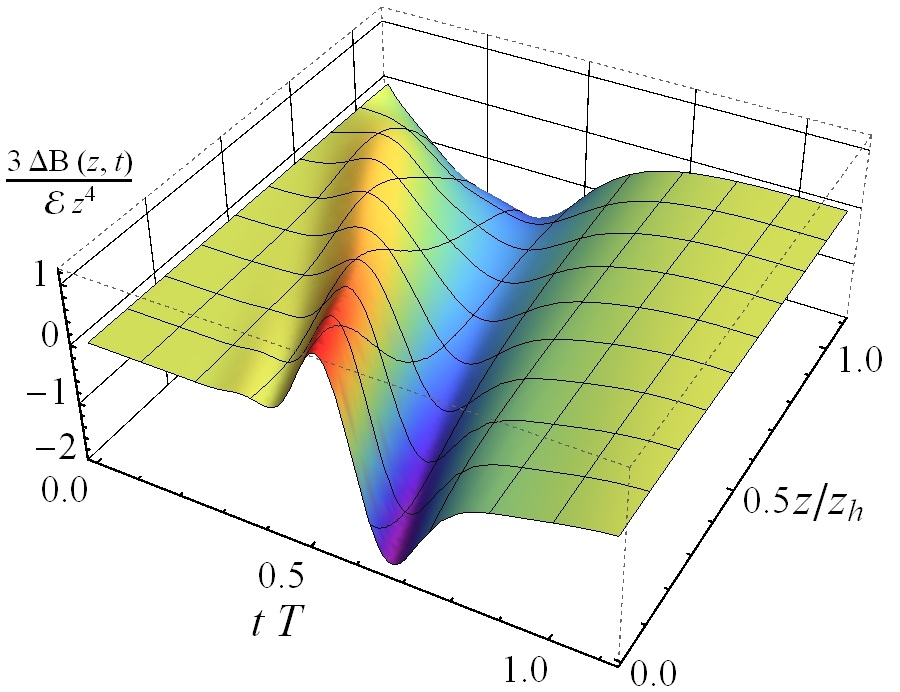}
\quad\quad \quad\\
 (a) & (b)\\ 
\end{tabular}
\caption{(a) Solution $B(t,z)$ (with $z\equiv 1/r$) obtained from the full Einstein's equations. The initial profile $B(t=0,z) = \frac{4}{5} (z/z_\mt{h})^4 \sin(8 z/z_\mt{h})$ is shown as a thick red curve. The thick blue curve shows $\deltap(t)/\ed$ as obtained from the full Einstein's equations. The thin magenta curve shows the value of $\deltap(t)/\ed$ as obtained from the linear approximation. (b) Difference between the full solution and the linear approximation.}
\label{fig:B3D}
\end{figure*} 
Fig.~\ref{fig:B3D}a shows the result from the full EEQs for a representative initial state. Fig.~\ref{fig:B3D}b shows the difference between the full solution and the LA for this state. The ratio in the overall scales of the plots, $2/10$, gives a rough estimate of the accuracy of the LA, namely 20\%, which  is remarkable given that the evolution is definitely far-from-equilibrium. This follows from the thick blue curve in Fig.~\ref{fig:B3D}a, which shows that the pressure anisotropy is almost an order of magnitude  larger than the energy density at some points during the evolution. 
 
\begin{figure}[t]
\centering{} 
\includegraphics[width=7cm]{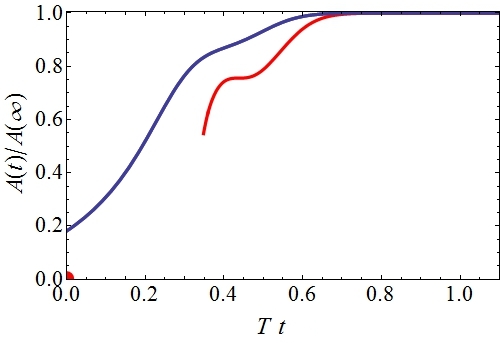}
\caption{Time evolution of the areas of the event (top, blue) and apparent (bottom, red) horizons for the initial state of Fig.~\ref{fig:B3D}a. The red dot at the origin signifies that  there is no apparent horizon for this state at the initial time. From that time until the start of the red curve there is no apparent horizon within the range of the radial coordinate covered by our grid, but there could be one at a deeper position.}
\label{entropy}
\end{figure}

Although there is no precise definition of entropy density far from equilibrium, it is interesting to examine the time evolution of the area densities of the event and apparent horizons, since these coincide with the entropy density in equilibrium. Fig.~\ref{entropy} shows that both of these quantities are larger at the end of the evolution than at the beginning, suggesting that entropy is indeed generated. Incidentally, note that  no entropy is produced in the LA, since $A$ and $\Sigma$ are unmodified. 

\begin{figure}[t]
\centering{}
\includegraphics[width=8.0cm]{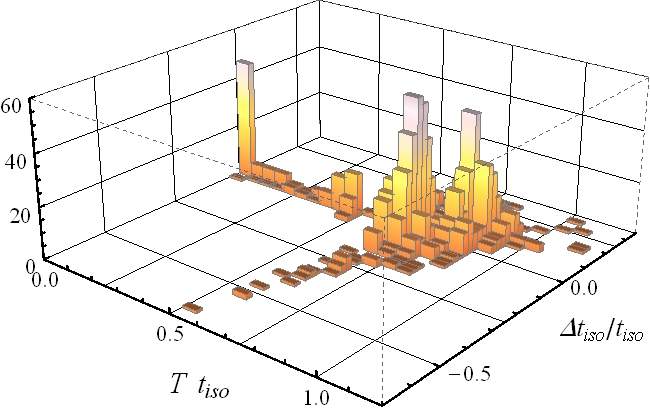}
\caption{Results for the isotropization times obtained from the full evolution of 1000 initial states, and for the differences between the full and the linearized evolution (normalized by the full isotropization time). The height of each bar indicates the number of states in  each bin. 
\label{fig:Tisos}}
\end{figure}

We define the isotropization time $\tiso$ as the time beyond which $\deltap(t)/\ed \leq 0.1$. Fig.~\ref{fig:Tisos} shows the isotropization times obtained from the full evolution of the 1000 initial profiles that we considered, as well as the differences between the values of these times as determined by the full EEQs and by the LA. We see that the LA works with a 20\% or better accuracy for most states and also that isotropization times are  $\tiso \lesssim 1/T$, with $T$ the final temperature.  

A quantitative analysis of the correlation between $\deltap/\ed$ and the produced entropy will be presented in \cite{preparation}. Suffice it to state here the qualitative trend: The larger the former, the larger the latter. In particular, if $\deltap/\ed\lesssim 1$, the entropy increase is fairly small ($\lesssim 10\%$).

\noindent
{{\bf 4. Discussion.}} Small perturbations around an equilibrated plasma can be described in linear-response theory. Equivalently, perturbations around the dual horizon can be described by linearizing EEQs around the equilibrium black hole solution. For a homogeneous but anisotropic perturbation one may expect the LA to be applicable whenever $\deltap/\ed \ll 1$. It is remarkable that, in a strongly coupled CFT with a gravity dual, the LA actually works relatively accurately well beyond this limit.  

In Fourier space, one may distinguish between hydrodynamic modes (HDMs), with dispersion relations $\omega(q)$ such that $\omega \to 0$ as $q\to 0$, and quasinormal modes (QNMs), for which  $\omega(0)\neq 0$. If the perturbation is anisotropic but homogeneous then the relaxation back to equilibrium involves exclusively the QNMs. In this sense the dynamics we have studied can be thought of as the far-from-equilibrium dynamics of the QNMs. If the perturbation is small then these modes evolve towards equilibrium linearly, independently of each other and on a time scale set by the imaginary parts of their frequencies. One could imagine extending the description to not-so-small perturbations by including non-linearities in the form of interactions between the QNMs, but naively one would expect this expansion to break down for order-one anisotropies. Instead, our results imply that, for a homogeneous, strongly coupled plasma with a gravity dual, the isotropization process is still reasonably well described by QNMs that evolve approximately linearly and independently of each other, even in the presence of large anisotropies (see \cite{Bantilan:2012vu} for related observations). This means that, just as in the near-equilibrium case, the relaxation towards equilibrium is characterized by a few quasinormal frequencies. In fact, we have verified that the time evolution of the stress tensor is well described by expanding and evolving $B$ in terms of a sufficient number of QNMs. In particular, a naive (under)estimate of the isotropization time can be obtained from the imaginary part of the lowest quasinormal frequency, which is $\mbox{Im} \, \omega_0 \simeq 8.5 T$. Our initial states have anisotropies of the order of $1 \lesssim \deltap /\ed \lesssim 20$, which  gives $0.27 \lesssim T \tiso \lesssim 0.62$. The reason why this is an underestimate is that the degree of anisotropy carried by each individual QNM can be much larger, typically as large as $\deltap /\ed \sim 500$, the total anisotropy being much smaller due to cancellations between different modes. With this value one gets $T \tiso \sim 1$.

It would be interesting to understand the reason behind the relative accuracy of the LA in our setup. This will certainly involve comparing the next-to-linear term in the expansion with the linear term. Yet,  demanding that the former is much smaller than the latter might yield too restrictive a condition, since subsequent terms in the expansion might partially cancel each other. 
 
We stress that the conclusions in this Letter hold for the purpose of computing the time evolution of the expectation value of the CFT stress tensor. Other observables, e.g.~those considered in \cite{Hubeny:2007xt,Chesler:2011ds}, may or may not be well described by the LA. This is under investigation.

Another interesting question is whether the LA may be applicable in less symmetric situations, for example in the absence of homogeneity, in which case HDMs will play an important role. Cases of particular interest include expanding plasmas, a simple example of which is a boost-invariant plasma. 
In this case the late-time solution is an expanding fluid with
temperature (at leading order in the hydrodynamic expansion)  $T(\tau) = \Lambda/(\Lambda\tau)^{1/3}$, with $\tau$ the proper time and $\Lambda$ a constant. The dual gravity solution is a black brane with a time-dependent horizon located at $r_\mt{h} (\tau) = \pi T(\tau)$ \cite{Janik:2005zt}. Thus it may be useful to linearize the EEQs around the late-time solution (possibly including known gradient corrections) 
if $\Lambda$ can be determined in terms of initial data defined at 
$\tau > 0$. This is currently under investigation. 

We close by stressing that the accuracy of the LA is a statement about the dynamics of black hole horizons. The LA is not expected to be applicable to the description of strong gravitational dynamics in the absence of horizons. In particular, it is not expected to be able to describe the formation of a horizon --- but it might be very useful in order to describe its subsequent evolution.  


\noindent
{{\bf Acknowledgements.}}
We thank H.~Bantilan, D.~Berenstein, J.~Casalderrey,  P.~Chesler, P.~de Forcrand, G.~Horowitz, T.~Jacobson, R.~Janik, J.~Noronha, F.~Pretorius,  J.~Santos, K.~Schalm, E.~Shuryak, C.~Sopuerta, U.~Sperhake and L.~Yaffe for discussions. 
WS thanks P.~Chesler for explaining the numerical method used in \cite{Chesler:2008hg}. MPH thanks the KITP at the University of California for hospitality during the completion of this project. This work was partially supported by the Polish Ministry of Science and Higher Education grant \emph{N N202 173539} and by the Foundation for Fundamental Research on Matter, which is part of the Netherlands Organization for Scientific Research. DM is supported by 2009-SGR-168, MEC FPA2010-20807-C02-01, MEC FPA2010-20807-C02-02, and CPAN CSD2007-00042 Consolider-Ingenio 2010. DT acknowledges support from CNPq.




\end{document}